\documentclass[aps,prl,twocolumn,showpacs,superscriptaddress,groupedaddress,nofootinbib]{revtex4}
\usepackage[dvips]{graphicx}
\usepackage{epsfig}
\usepackage{subfigure}
\usepackage{float}
\usepackage{color}
\usepackage{ucs}
\usepackage[utf8x]{inputenc}
\usepackage{soul}
\usepackage{xr-hyper}
\begin{document}

\title{Infrared Peak-Splitting from phonon localization in Solid Hydrogen}
\date{\today} \author{Ioan B. Magdău} \author{Graeme J. Ackland}
\affiliation{School of Physics and Centre for Science at Extreme
  Conditions, University of Edinburgh, Edinburgh EH9 3JZ, UK.}

\pacs{61.50.Ah,  63.20.dk 63.50.Gh 78.30.Ly}

\begin{abstract}
We show that the isotope effect leads to a completely different
spectroscopic signal in hydrogen-deuterium mixtures, compared to pure
elements that have the same crystal structure.  This is particularly
true for molecular vibrations, which are the main source of
information about the structure of high-pressure hydrogen. Mass
disorder breaks translational symmetry, meaning that vibrations are
localized almost to single molecules, and are not zone-center phonons.
In mixtures, each observable IR peak corresponds to a collection of
many such molecular vibrations, which have a distribution of
frequencies depending on local environment.
Furthermore discrete groups of environments cause the peaks to split.
We illustrate this issue by considering the IR spectrum of the high
pressure Phase III structure of hydrogen, recently interpreted as
showing novel phases in isotopic mixtures.  We calculate the IR
spectrum of hydrogen/deuterium mixtures in the $C2/c$ and $Cmca$-12
structures, showing that isotopic disorder gives rise to mode
localization of the high frequency vibrons. The local coordination of
the molecules leads to discrete IR peaks.  The spread of frequencies
is strongly enhanced with pressure, such that more peaks become
resolvable at higher pressures, in agreement with the recent
measurements.
\end{abstract}

\maketitle

The phase diagram of hydrogen is of intense interest at present as the
search for crystalline metallic hydrogen hots
up\cite{dias2016new,dalladay2016evidence,howie2015raman,howie2014phonon,eremets2016low}. Critically,
metallic hydrogen must not only be made\cite{SilveraMetalScience}, but
an unambiguous experimental signal to prove its existence is needed.
It is therefore central to the endeavour that the expected spectroscopic 
signature is properly understood.
  
Due to the poor
X-ray scattering and small sample sizes, precision crystallography of
high pressure hydrogen has proved impossible and most information is
gleaned from Raman and Infrared (IR) spectroscopy.  This measures the
vibrational frequencies within the crystal, and cannot determine
crystal structure.  However, it often provides enough information to
distinguish between theoretically produced candidate structures. For
this combined calculation-experiment solution of crystal structure 
to work, it is essential that the calculated spectra are well understood.

Because of the mass difference between the isotopes, the spectroscopic
signal in hydrogen and deuterium is different. For harmonic phonons,
at the same atomic volume this gives a simple frequency rescaling of
$\sqrt{2}$. Variations from this give information about anharmonic
behaviour, which will normally be larger for the lighter isotope.  But
there are practical difficulties in ``equal volume'' comparisons.  For
a given volume, the calculated pressure and free energy depend on zero
point effects, which can be different in D$_2$ and H$_2$ by up to
10 GPa\cite{ackland2015appraisal,mcminis2015molecular,drummond2015quantum}.  Volume is extremely difficult to
measure experimentally, while the ``diamond edge'' pressure scale
has been twice recalibrated by up to 20 GPa in the last ten
years\cite{akahama2010calibration}.

Although the H$_2$ and D$_2$ spectra are related by a simple scaling,
the spectrum of hydrogen deuteride (HD) can be far more complicated.
This is mainly because, once the reaction $2HD\Leftrightarrow H_2 +
D_2$ reaches equilibrium, the solid contains three molecular types.
Molecular modes such as vibrons and rotons then split into three,
while lattice modes such as layer vibration adopt an average
frequency.  Furthermore, because of mass-disorder, Bloch's theorem no
longer applies and the difference in masses can drive mode
localization, provided the frequency is high enough
\cite{anderson1958absence,monthus2010anderson}.  These features mean
that the spectroscopic signal from hydrogen-deuterium mixtures,
whether Raman on IR, can look very different from the single element
even if the underlying structure is identical.  The importance of high
frequency in allowing localization means that it is the vibron spectra
where this is most pronounced\cite{monthus2010anderson,howie2014phonon}.

Phase III of elemental H$_2$ and D$_2$   has a particularly strong and simple
IR signature\cite{hemley1988phase,lorenzana1989evidence,goncharov2001spectroscopic,loubeyre2002optical,zha2012synchrotron,goncharov2013hydrogen}, so it was surprising when very
complex spectra were found in recent experiments starting with HD at low
temperatures\cite{dias2016new}.  At low pressures only
the HD infrared peak was observed, corresponding to Phase II.  Upon
entering phase III, an H$_2$ peak was observed, but no D$_2$. 
Subsequently two new phases were reported, based on the
infrared spectra. The first transition to a new phase
IV* happened at 200 GPa where both HD and H$_2$ peak split in two and
two new D$_2$ vibron peaks appeared. The second transition around 250
GPa to a phase called HD-PRE was identified by further splitting in the
HD vibron.  

Different phase behaviour in isotopic mixtures from pure elements is
extremely difficult to understand with equilibrium thermodynamics.  The
electronic structure in the Born-Oppenheimer approximation is the same
in each case, so isotopic differences are due to vibrational effects,
notably zero point energy, which depends on mass
\footnote{We disregard the possibility that all previous experiments
  on H/D were somehow trapped in an para/ortho ground state, and that
  those different nuclear states somehow lead to the same energetics
  which are, in turn, different from the energetics of HD which has no
  ortho/para states.}.  Hydrogen and deuterium exhibit the same phase
sequence, so different HD phases would require that, as mean molecular
mass increases from 1, to 1.5 to 2, the structure is first
destabilized then restabilized by increasing mass.  This observation
is sufficiently unexpected as to merit further explanation.

Density functional theory (DFT)  calculations using the PBE functional have
produced a range of possible candidate structures for hydrogen under
these conditions, with good agreement for Raman and IR spectra of
phases III and
IV in pure isotopes\cite{clark2005first,perdew1996generalized,pickard2007structure,pickard2012density,lebegue2012semimetallic,magdau2013identification,monserrat2016hexagonal}.
In calculations with static ions, DFT exhibits no
isotope effects: these enter only through the vibrational behaviour, mainly
zero-point effects.  Spectroscopy
probes this vibrational behaviour, so isotopic differences can be studied using lattice dynamics.  Here we
calculate the expected IR spectroscopic signature from candidate
structures above 150 GPa and below room temperature.

\begin{figure}[h!]
  \includegraphics[width=1\textwidth]{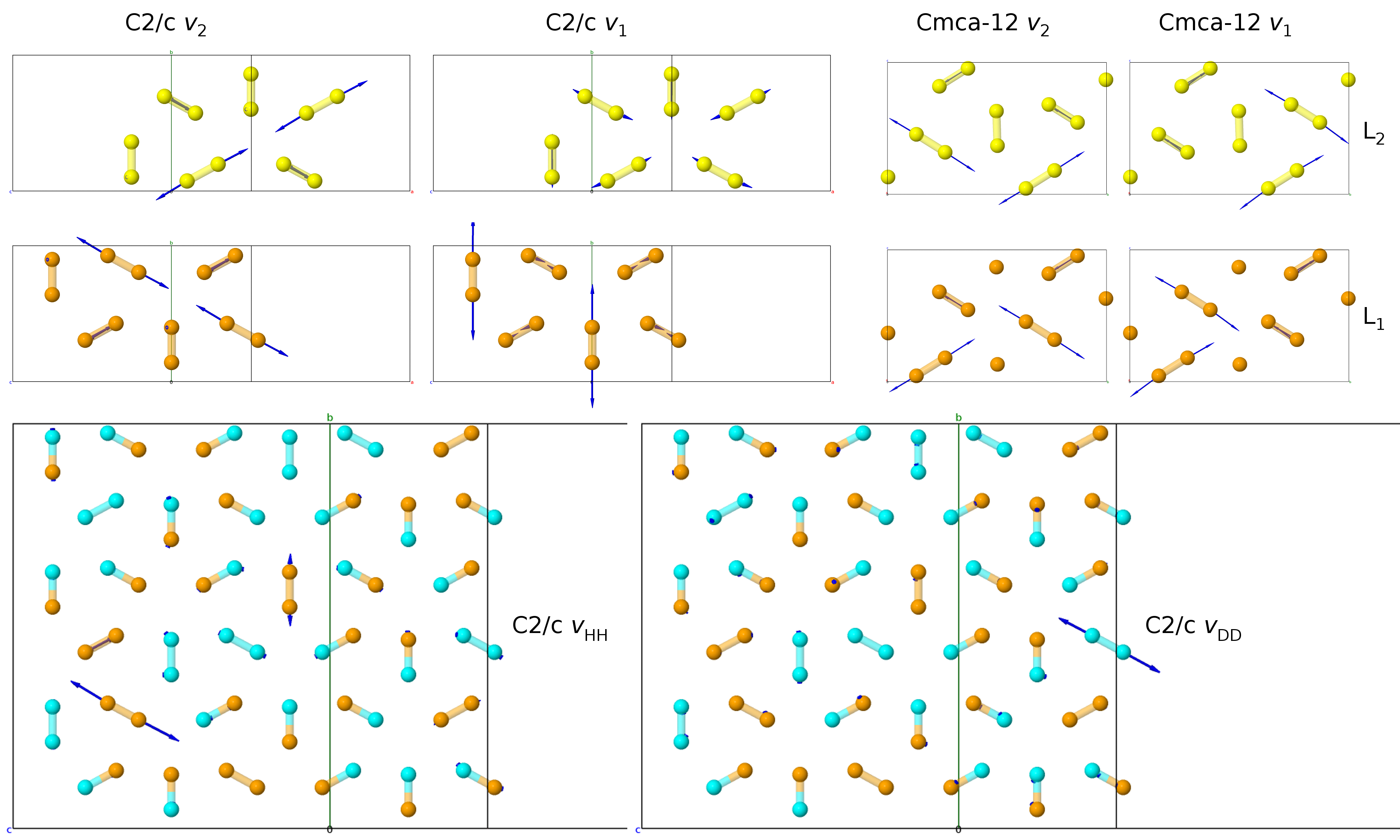}
  \caption{
    \label{PhononsVisual}
    Upper panels show the two most intense IR-active vibron modes at
    250 GPa in pure hydrogen, assuming (a) $C2/c$ and (b) $Cmca$-12
    structures (24 atoms). The unit cells comprise two layers, shown
    separately for clarity ($L_1$ - orange and $L_2$ - yellow). (c)
    Displacements corresponding to IR-active vibrons from
    $C2/c$ cell (288 atoms) of disordered mixture at 250
    GPa. Orange: H; cyan: D.
    Isotopic symmetry breaking means that all modes obtain some
    IR-activity. These highly localized vibron modes shown are
    representative examples for the spectrum generated by mixtures.}
\end{figure}

\begin{figure}[h!]
  \includegraphics[width=0.5\textwidth]{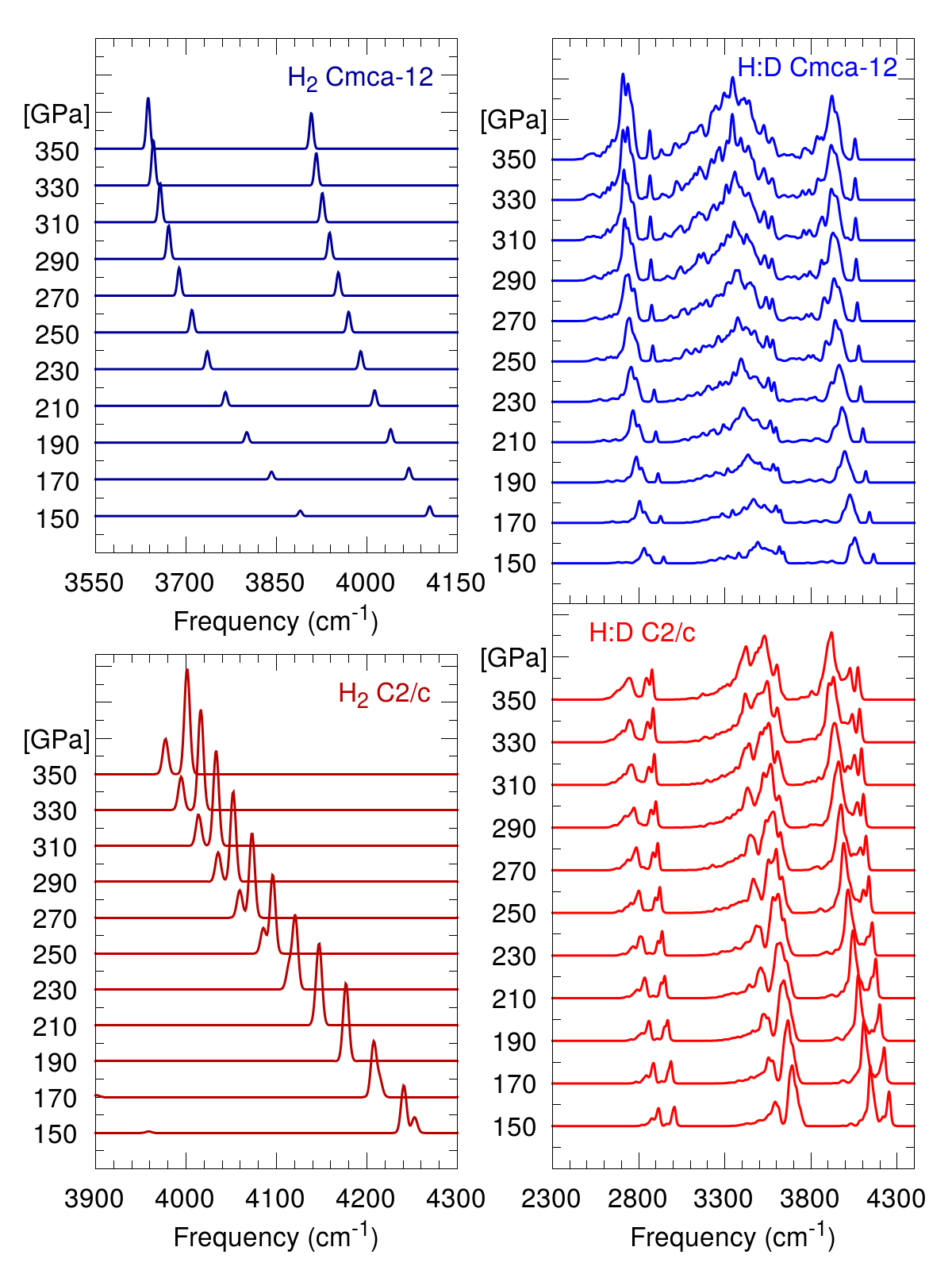}
  \caption{
    \label{SpectraVsPres}
   Stacked plot of simulated vibronic infrared signal at different
   pressures (shown on y-axis) for the two candidate structures $C2/c$
   and $Cmca$-12.  Intensities are calculated so spectra at different
   pressure points are comparable.  
   On the left pure hydrogen as (a) $Cmca$-12 (b) $C2/c$. 
   Normal-mode linewidths narrower than experimental
   resolution\cite{zha2012synchrotron} are chosen to emphasize the
   impossibility of resolving the two $C2/c$ modes below 230 GPa.  On
   the right we show the data for 50:50 hydrogen-deuterium mixtures  
   (c) $Cmca$-12 (d) $C2/c$.
   Comparison in Fig\ref{CompToExp} (and SM) shows that $C2/c$ is a
   better candidate than $Cmca$-12. At low pressures $C2/c$ has six
   distinct peaks, two for each molecular type (H$_2$, HD, D$_2$). The
   two peaks per molecule type here are not connected to the distinct
   peaks in the pure hydrogen spectra, but rather a signature of local
   molecular environments. This splitting is an indirect indication of
   mode localization, which confers some IR activity to all modes
   \cite{azadi2013fate}.}
\end{figure}

We calculated the pressure evolution of the infrared spectra for 50:50
hydrogen-deuterium mixtures at the equilibrium concentrations of 25\%
H$_2$, D$_2$ and 50\% HD.  According to
now-routine calculations
the energetically-favored crystal  is one of a number of similar 
structures such as
$C2/c$ or  $P6_122$ which can be viewed as atomic layers\cite{pickard2012density,monserrat2016hexagonal}, or as close-packed
molecules\cite{magdau2015simple}.
We choose to represent these phases
with $C2/c$ and for contrast, we also consider the $Cmca$-12
structure, predicted to be stable at higher pressures  (Fig.\ref{PhononsVisual}.  ).  We calculated the
full Raman spectra, but we concentrate on the vibron modes because they can
be directly compared with experiment (Fig.\ref{SpectraVsPres}) and
because the low frequency modes do not show localization
effects\cite{monthus2010anderson}.

$C2/c$ has two strongly IR-active vibron modes for pure hydrogen,
which reduce in frequency with increased pressure. This unusual
softening can be traced to the weakening of the covalent bond, which
lengthens and loses charge as pressure increases.  A cursory
inspection of the full pressure dependence for $C2/c$
(Fig.\ref{SpectraVsPres}) shows two peaks at the onset of phase III
(150 GPa), a single peak in the range 170-230 GPa, and two peaks above
250 GPa.  Based only on these results, it would be easy to draw a
mistaken conclusion that this peak ``splitting'' signified a phase
transition: but since all calculations are in the same phase this can
be ruled out - it is simply the effect of different pressure
dependencies of the mode frequencies.  To avoid this type of
confusion, in this letter we draw a distinction between ``peaks'' -
the spectroscopic observables and ``modes'' - their calculated cause.

\begin{figure}[h!]
  \includegraphics[width=0.45\textwidth]{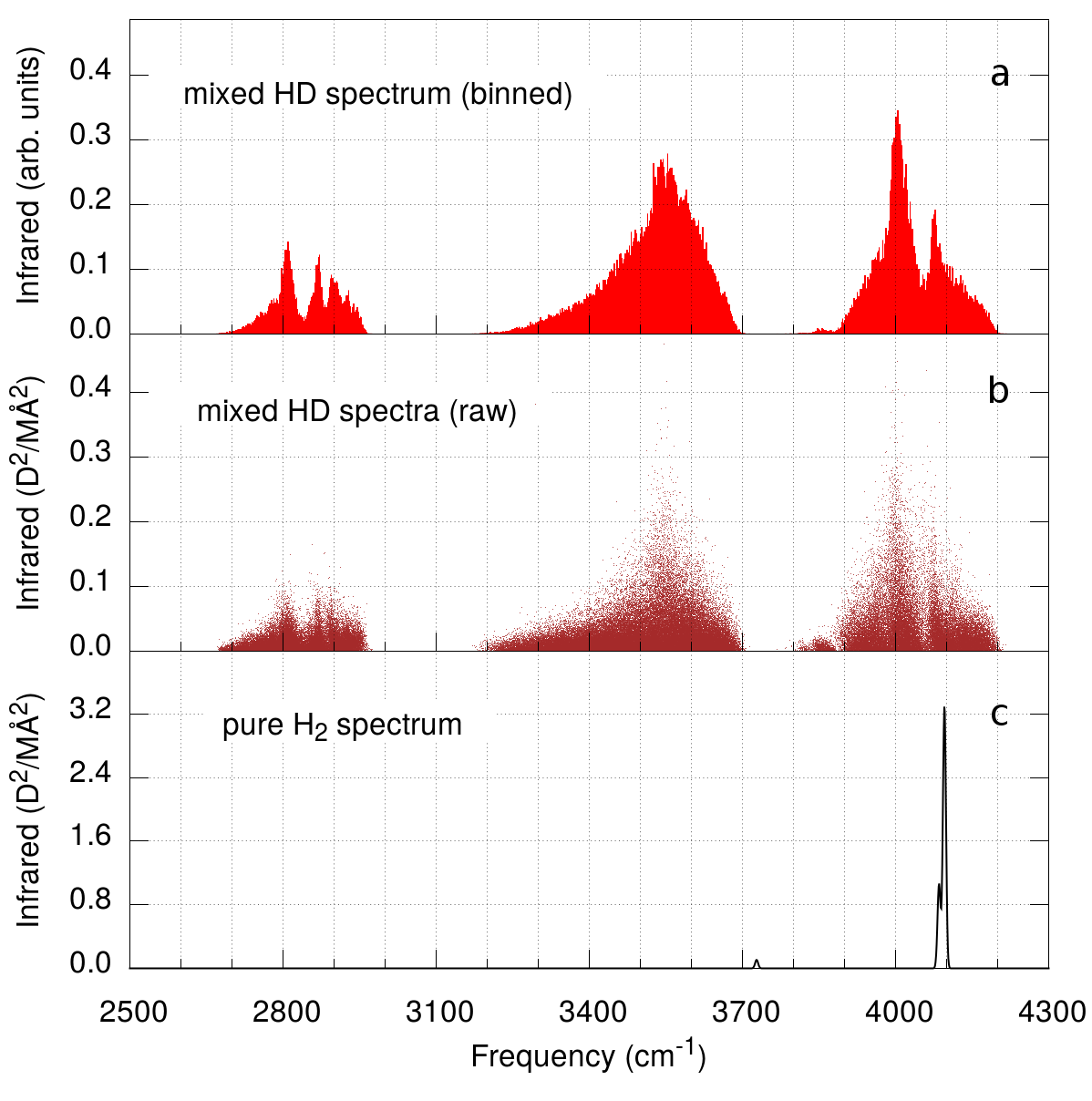}
  \caption{
    \label{BuildingSpectraMethod}
    Method used to calculate spectra in mixtures at 250 GPa from 288
    atom calculations, explained in detail in SM\cite{SM}.  (a)
    mixture data binned up in a histogram for comparison with
    experiment.  (b) dot for the intensity of each of 144000 localized
    modes generated from 1000 random samples of hydrogen-deuterium
    mixtures.  (c) IR spectrum of pure hydrogen $C2/c$.}
\end{figure}

Lattice dynamics in mixtures is considerably more involved than in
pure elements, and we have developed sophisticated methods to tackle
the problem using extended supercells (See
Fig.\ref{BuildingSpectraMethod}\cite{SM,howie2014phonon}). The
difficulty arises because the disorder breaks translational symmetry so
that crystal momentum, $k$, is no longer a good quantum number and the
$\Gamma$-point selection rule $k=0$ becomes meaningless.  The twofold
D/H mass difference is sufficient to cause the vibron modes to 
become localized.  This broken symmetry confers IR activity on all modes.  
Comparing
Fig.\ref{BuildingSpectraMethod}b and \ref{BuildingSpectraMethod}c
shows that the IR activity of any individual mode in the mixture is much
lower than for the $\Gamma$-point phonon in $H_2$. So whereas the pure H$_2$
spectrum has peaks corresponding to a few unique normal modes, the
peaks in the mixtures correspond to many different localized modes
(c/f Fig.\ref{PhononsVisual}).  The localised modes form three groups
based on H$_2$, HD, and D$_2$ vibrations, with each group split
further due to different local environments.

Coupling between molecules causes dispersion in vibron energies.
Typically, the antisymmetric IR-active modes are shifted to higher
frequencies and the symmetric Raman-modes to lower frequencies.  In HD
mixtures, the coupling is weaker than in pure hydrogen because $H_2$
$HD$ and $D_2$ have different resonant frequencies.  So at an
equivalent pressure the dispersion-like effects are smaller.
Consequently strongly IR-active $H_2$-type modes can be expected to
have lower frequencies than in pure $H_2$, while Raman frequencies for
the equivalent modes in mixtures will tend to be higher. Because all
modes have some IR activity, the peaks are skew with a low-frequency
tail.
  
The H$_2$ IR vibron frequency is sometimes used experimentally to
measure pressure.  The different dispersion between H$_2$-like peaks in mixtures
and pure elements makes this approach unreliable.

In our isotopic mixture calculations, at low pressures, we find that the modes
cluster to produce six clear distinct vibron peaks in $C2/c$, two for
each molecular species H$_2$, HD and D$_2$. These modes are typically
localized on one molecular type (Fig. \ref{PhononsVisual}).  At
higher pressures (above 230 GPa), we find the remarkable result that
all the peaks split further, but most notably the HD signal splits
into three peaks.  The qualitative difference from the pure elements
arises from the existence of a finite number of well-defined
environments in which the molecule can find itself. This type of
splitting is also observed in the $Cmca$-12 structure.

The central result from the calculations is that the different
appearance of the IR spectra in isotopic mixtures compared to pure
elements is a consequence of mode localization rather the different
crystal structures.

We now compare our results to the recent experiments reported by Dias
et al.  For convenience the main data from that paper are reproduced
in the supplemental materials.  Consistent with previous studies of pure H$_2$ and D$_2$
and free energy calculation assuming the $C2/c$ structure they report
the onset of phase III around 150 GPa.  The observed IR spectrum (See
SM\cite{SM,dias2016new} is consistent with our lattice
dynamics (Fig.\ref{PhononsVisual}c) assuming mass scaling appropriate for
HD molecules.

At around 200 GPa, Dias et al report a change in the IR spectra with
new peaks appearing at frequencies consistent with Phase III of H$_2$
and D$_2$.  Each of these peaks is further split, so they label this
Phase HD-IV*, while noting that its phase lines are very different
to the previously-named Phase IV and IV'.

We presume that this change is associated with the conversion of
$2HD\Leftrightarrow H_2 + D_2$.  The equilibrium state has H and D
distributed randomly between molecules, consistent with calculations
showing negligible dependence of binding energy on isotope
ordering\cite{howie2014phonon}.  After the transition from II to III,
the HD dissociation might occur by lattice rebonding.  Previous
molecular dynamics calculations in similar layers of Phase IV have
shown that the isotope-disordered equilibrium can be reached by
solid-state molecular rebonding effects on the picosecond timescale,
albeit at higher temperatures\cite{liu2012room,magdau2014high}, and
that the dissociation rate increases exponentially with pressure as
the bond weakens.

\begin{figure}[h!]
  \includegraphics[width=1\textwidth]{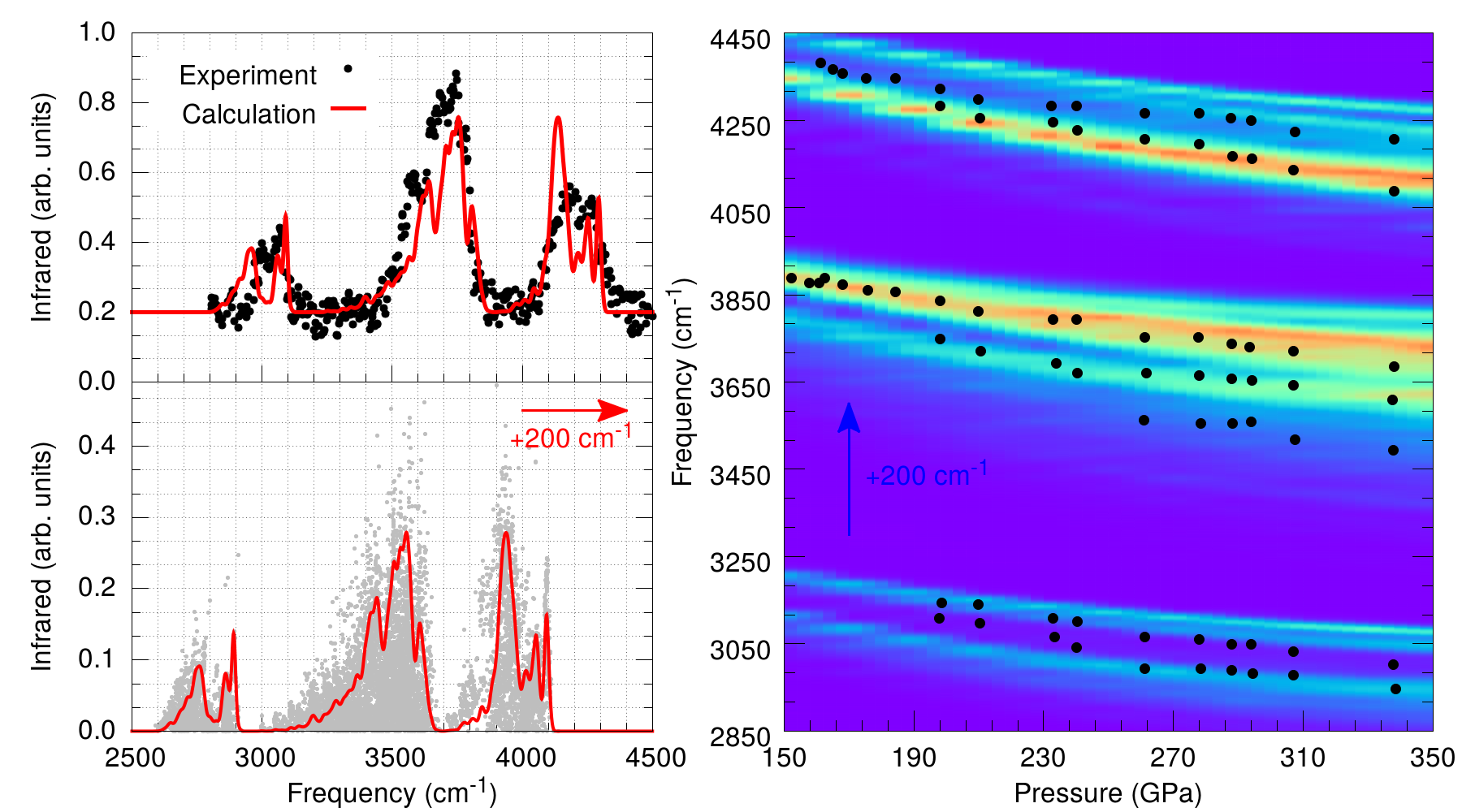}
  \caption{
    \label{CompToExp}
    Comparison of calculated IR in $C2/c$ mixtures (see also
    Fig.\ref{SpectraVsPres}) to the recent raw experimental data
    \cite{dias2016new}. We have removed the Gaussian fits made 
    to the data by Dias {\it et al} since this approach is invalid 
    to describe localised modes.  Theoretically calculated spectra are
    systematically some $200$ cm$^{-1}$ softer than experimental
    spectra, due to a combination of errors including 
neglect of
    temperature, zero point effects in calculated pressure,
    calibration of the experimental pressure and choice of
    exchange-correlation functional
    \cite{ackland2015appraisal,akahama2010calibration,azadi2013fate}. (a) $C2/c$ spectra at 300 GPa from 3000 randomized samples: gray
    dots are individual modes, red line is a histogram. (b)
    Comparison of the calculations to the digitized experimental
    data at 307 GPa\cite{dias2016new,rohatgi2011webplotdigitizer}. (c) Color mapshowing the calcaulated  pressure dependence of the IR
    intensity compared to the equivalent experimental
    data\cite{dias2016new}.}
\end{figure}

To compare our calculated results with experiment, we calculated
phonon frequencies and IR intensities at a range of pressures from 3000
isotopically-disordered samples based on the $C2/c$ structure. In Fig.\ref{CompToExp} we compare experimental
data points \cite{dias2016new} and calculations without making any assumption
about the number of peaks in either case.  Our calculations
of mixtures in the $C2/c$ phase produce peak splitting in
excellent agreement with this data: $Cmca−12$ is significantly worse.  
Similar analysis has been done
across a range of pressures, and by tracing the evolution it is possible
to pick out persistent features. Unlike the phonons of pure hydrogen,
these peaks all have similar pressure dependence.  This surprising
result can be traced back to the localisation of the modes, and the
splitting being due to the nearby environment rather than the long
range symmetry. As pressure increases stronger coupling between
molecules both broadens the splitting and equalises the IR activity of
the modes, making it possible to resolve the distinct local
environments as seen in the HD-PRE phase.  The important point is that
the improved ability of resolve multiple peaks with increasing
pressure does not correspond to a breaking of symmetry or to a
structural phase transition.

Thus we propose that observed phases III, IV* and HD-PRE, all have the
same crystal structure, probably $C2/c$\cite{pickard2012density} or a
closely related structure\cite{monserrat2016hexagonal}.  The observed
changes in the IR spectrum are due to isotopic effects rather than
structural change.  Free-energy calculations imply that III-IV*
involves an isostructural $2HD\rightarrow H_2+D_2$ equilibration
reaction and that only the structure with disordered H and D atoms should appear on the equilibrium phase diagram.

As the thermodynamic equilibrium phase IV* is approached, D$_2$
molecules form and a peak corresponding to D$_2$ vibrons is
observed\footnote{we note that the experimental sample had a slight
  excess of hydrogen over deuterium}.  The splitting of the IR vibron peaks in
  IV* is due to different modes and local environments in
  isotope-disordered $C2/c$, as opposed to the two-layer structure in
  Phase IV, IV' and V.

We have shown that due to mode localisation the Raman and Infrared
spectra of hydrogen-deuterium mixtures are a lot more complex than
those of pure isotopes, even for the same crystal structures.
Moreover, under pressure the dispersion of phonon bands increases,
such that vibron frequencies move further apart.  So our main message
is that the observation of different numbers of well-resolved peaks
does not necessarily indicate a different crystal structure.  This 
applies both to a single system under pressure, where modes have
different pressure dependencies (Fig. \ref{SpectraVsPres}) and when
comparing spectra of disordered mixtures to those with a single
molecular type.

These effects are sufficient to explain the multiple peaks observed in
recent IR data without recourse to new crystal
structures\cite{dias2016new}.  There is no reason to suppose that the
equilibrium phase diagram for hydrogen-deuterium mixtures contains any
phases other than those observed in pure hydrogen.

Thanks to the richness of the spectra, spectroscopic measurement with
the resolution reported by Dias et al on mixtures provides important
data which cannot be obtained from pure H$_2$ or D$_2$.  Although
spectroscopy does not provide a conclusive crystal structure
determination, in combination with calculation $Cmca-12$
can be ruled out and a layered structure such as
with molecules arranged close to hcp such as $C2/c$ remains a favored
candidate for Phase III\cite{magdau2015simple}.

Very recent work reporting cold metallic hydrogen remains
controversial\cite{SilveraMetalScience}, there is no indication about
the crystal structure, or even whether the sample is crystalline. This
can be addressed with spectroscopy, but as yet no data is available.
When it comes, the resolution is likely to be poor and interpretation
difficult, and to relate it to candidate crystal structures it will be
essential that it is analysed correctly as we have described.  Our
work here shows that studies of isotopic mixtures will provide
additional information not available from pure hydrogen or deuterium,
which will be crucial in finally deciphering how the long-sought
Wigner-Huntingdon transition occurs.

\acknowledgments{We thank E.Gregoryanz, I.Silvera, R. P. Dias, and O
  Noked for useful discussions about experimental details and
  providing their raw data. We thank M.Martinez-Canales, C.J.Pickard,
  B.Monserrat for insightful comments about the calculations.  We
  thank the Archer computing service at EPCC (EPSRC grant K01465X and
  a studentship).  GJA acknowledges support from the ERC fellowship
  ``Hecate'' and a Royal Society Wolfson fellowship.}

\onecolumngrid

\twocolumngrid

\onecolumngrid

\twocolumngrid

\bibliographystyle{apsrev}
\bibliography{Refs}

\end{document}